\documentclass[12pt]{article}
\usepackage[cp1251]{inputenc}
\usepackage{amsfonts}
\usepackage{latexsym}
\begin{document}

\begin{center}
{\Large \bf Consistent ADD scenario with\\ stabilized extra dimension} \\

\vspace{4mm}
%                      author/address

Mikhail N. Smolyakov\\ \vspace{0.5cm} Skobeltsyn Institute of
Nuclear Physics, Moscow State University
\\ 119992 Moscow, Russia \\
\end{center}

\begin{abstract}
A model with one compact extra dimension and a scalar field of
Brans-Dicke type in the bulk is discussed. It describes two branes
with non-zero tension embedded into the space-time with flat
background. This setup allows one to use a very simple method for
stabilization of the size of extra dimension. It appears that the
four-dimensional Planck mass is expressed only through parameters
of the scalar field potentials on the branes.
\end{abstract}

\section{Introduction}
\label{intro} Nowadays models with extra space-time dimensions are
widely discussed in the literature. A lot of problems are
addressed with the help of such models, in particular, the
hierarchy problem. In the latter case the most known models are
the ADD scenario \cite{ADD,Ant} and the Randall-Sundrum model with
two branes (usually abbreviated as RS1 model) \cite{RS1} (both
with compact extra space-like dimensions). The ADD scenario seems
to be simpler, but the brane tension is not taken into
consideration in this model. This problem was solved within the
framework of the RS1 model, nevertheless this model appeared to be
much more complicated. Both models were widely discussed in the
literature, see reviews \cite{Rub01,Kubyshin}.

Here we propose a model, which unifies some features of the models
mentioned above and can be treated as a "consistent ADD". It
describes branes with tension in the five-dimensional flat bulk
and allows one to use simple method for stabilization of extra
dimension's size.

\section{Description of the model}
\label{sec:1} Thus, we consider a space-time with one compact
extra space-like dimension. Let us denote the coordinates by $ \{
x^M\} \equiv \{x^{\mu},y\}$, $M= 0,1,2,3,4, \, \mu=0,1,2,3$, the
coordinate $x^4 \equiv y$ parameterizing the fifth dimension. As
in the RS1 model, it forms the orbifold $S^{1}/Z_{2}$, which is
realised as the circle of the circumference $2L$ with points $y$
and $-y$ identified. It is evident that the metric $g_{MN}$
satisfies the corresponding orbifold symmetry conditions. The
branes are located at the fixed points of the orbifold, $y=0$ and
$y=L$.

The action of the model is chosen to be
\begin{eqnarray}\nonumber
S&=&\int \Phi(x,y) R\sqrt{-g}\,d^{4}xdy+\\ \nonumber &+&
\int_{y=0}\left(\lambda_{1}-\gamma_{1}\left[\Phi^{2}(x,0)-v^{2}_{1}\right]^{2}\right)\sqrt{-\tilde
g}\,d^{4}x+\\ \label{action}
&+&\int_{y=L}\left(\lambda_{2}-\gamma_{2}\left[\Phi^{2}(x,L)-v^{2}_{2}\right]^{2}\right)\sqrt{-\tilde
g}\,d^{4}x
\end{eqnarray}
where $\Phi(x,y)$ is a five-dimensional scalar field (such that
$\Phi(x,-y)=\Phi(x,y)$), parameters $\lambda_{1}$, $\lambda_{2}$,
$\gamma_{1}$, $\gamma_{2}$, $v_{1}$ and $v_{2}$ describe the
scalar field potentials on the branes (or simply the brane
tensions), $\tilde g_{\mu\nu}$ is the induced metric on the
branes. The first term in (\ref{action}) corresponds to the
five-dimensional Brans-Dicke theory with the Brans-Dicke parameter
equal to zero, whereas the last two terms correspond to the
contribution of the branes, where the Higgs-like potentials for
the field $\Phi(x,y)$ are added (surely one can choose another
type of potentials). It is necessary to note that the
five-dimensional theory with Brans-Dicke field is widely discussed
with relation to cosmology, see \cite{Mendes,Arik} and references
therein.

The equations, following from action (\ref{action}), have the form
(see, for example, \cite{Thorn})
\begin{eqnarray}\nonumber
& &
\Phi\left(R_{MN}-\frac{1}{2}g_{MN}R\right)-\left(\nabla_{M}\nabla_{N}\Phi-
g_{MN}g^{AB}\nabla_{A}\nabla_{B}\Phi\right)-
\delta^{\mu}_M\delta^{\nu}_N \times\\ \label{ee} &\times
&\left(\frac{\sqrt{-\tilde g}}{\sqrt{-g}}\tilde
g_{\mu\nu}\frac{\lambda_{1}-\gamma_{1}\left[\Phi^{2}-v^{2}_{1}\right]^{2}}{2}\,\delta(y)+\frac{\sqrt{-\tilde
g}}{\sqrt{-g}}\tilde
g_{\mu\nu}\frac{\lambda_{2}-\gamma_{2}\left[\Phi^{2}-v^{2}_{2}\right]^{2}}{2}\,\delta(y-L)\right)=0
\end{eqnarray}
(the Einstein equations), where $\nabla_{M}$ is the covariant
derivative with respect to the metric $g_{MN}$, and
\begin{eqnarray}\label{scalar}
R-\frac{\sqrt{-\tilde
g}}{\sqrt{-g}}4\gamma_{1}\Phi\left(\Phi^{2}-v^{2}_{1}\right)\,\delta(y)-\frac{\sqrt{-\tilde
g}}{\sqrt{-g}}
4\gamma_{2}\Phi\left(\Phi^{2}-v^{2}_{2}\right)\,\delta(y-L)=0
\end{eqnarray}
(the equation for the scalar field).

We are going to find a solution of equations (\ref{ee}),
(\ref{scalar}) with $g_{MN}=diag(-1,1,1,1,1)$, i.e. with the flat
five-dimensional background metric. Let us suppose that background
solution for the field $\Phi(x,y)$ does not depend on the
four-dimensional coordinates $x$. In this case the only
non-trivial component of (\ref{ee}) takes the form
\begin{equation}
g_{\mu\nu}\Phi''-\left(\tilde
g_{\mu\nu}\frac{\lambda_{1}}{2}\,\delta(y)+\tilde
g_{\mu\nu}\frac{\lambda_{2}}{2}\,\delta(y-L)\right)=0,
\end{equation}
where $\Phi''=\frac{d^2\Phi}{dy^2}$ and equation (\ref{scalar}) is
taken into account. If
\begin{equation}\label{finetune}
\lambda_{1}=-\lambda_{2}=\lambda,
\end{equation}
we get
\begin{equation}
\Phi(y)=\frac{\lambda}{4}|y|+C,
\end{equation}
where $C$ is a constant and the orbifold symmetry conditions are
taken into account. From equation (\ref{scalar}) one can get
\begin{eqnarray}
\Phi(y)=\Phi_{0}=\frac{\lambda}{4}|y|+v_{1},\\ \label{L}
L=\frac{4}{\lambda}\left(v_{2}-v_{1}\right).
\end{eqnarray}
We see that the size $L$ of extra dimension is stabilized (by
equation (\ref{scalar})). It is necessary to note that in the case
$\lambda>0$ physically relevant result can be obtained only if
$v_{2}>v_{1}>0$.

Now let us find effective four-dimensional Planck mass on the
branes. To this end it is necessary to obtain the wave-function of
the massless (from the four-dimensional point of view) tensor mode
of $h_{\mu\nu}$, which is the fluctuation of $\mu\nu$-component of
the metric. Linearizing $\mu\nu$-component of equation (\ref{ee})
and dropping the scalar degrees of freedom, one can get the
following equation (in the transverse-traceless gauge for the
field $h_{\mu\nu}$):
\begin{equation}\label{lineq}
\Phi_{0}\Box h_{\mu\nu}+\Phi_{0} h''_{\mu\nu}+\Phi_{0}'
h'_{\mu\nu}=0.
\end{equation}
The third term of this equation comes from the term
$\nabla_{\mu}\nabla_{\nu}\Phi$ in (\ref{ee}). Here
$\Box=\eta^{\mu\nu}\partial_{\mu}\partial_{\nu}$. The field
$h_{\mu\nu}$ can be represented as a sum
\begin{equation}\label{summ}
h_{\mu\nu}(x,y)=\sum_{n} h^{n}_{\mu\nu}(x)\Psi_{n}(y),
\end{equation}
where $\Box h^{n}_{\mu\nu}=m_{n}^{2}h^{n}_{\mu\nu}$ and
$\Psi_{n}(y)$ are the wave functions which correspond to the
four-dimensional masses $m_{n}$. Equation (\ref{lineq}) takes the
form
\begin{equation}\label{lineq1}
\Phi_{0}m_{n}^{2}\Psi_{n} +\Phi_{0} \Psi''_{n}+\Phi_{0}'
\Psi'_{n}=0,
\end{equation}
which is equivalent to
\begin{equation}\label{lineq-1}
\Phi_{0}m_{n}^{2}\Psi_{n} +\left(\Phi_{0} \Psi'_{n}\right)'=0,
\end{equation}
where the differential operator has the self-adjoint form. It is
not difficult to show that
\begin{equation}
\int_{-L}^{L}\Phi_{0}(y)\Psi_{n}(y)\Psi_{k}(y)\sim\delta_{nk},
\end{equation}
i.e. the eigenfunctions of different modes are orthogonal. It is
evident that $\Psi_{0}(y)=const$. Other eigenfunctions of equation
(\ref{lineq1}) have the form
\begin{equation}\label{e-func}
\Psi_{n}(y)=C_{1}J_{0}\left(m_{n}\left[|y|+\frac{4v_{1}}{\lambda}\right]\right)+
C_{2}N_{0}\left(m_{n}\left[|y|+\frac{4v_{1}}{\lambda}\right]\right),
\end{equation}
where $C_{1}$ and $C_{2}$ are constants, $J_{0}$ and $N_{0}$ are
Bessel and Neumann functions. Substituting (\ref{e-func}) into
(\ref{lineq1}) and collecting terms with delta-functions (coming
from $|y|$), one can get the conditions on the "boundaries"\ $y=0$
and $y=L$, which define the mass spectrum of the theory and the
relation between $C_{1}$ and $C_{2}$ and look like
\begin{eqnarray}\label{boundone}
\Psi'_{n}(0)=0,\\ \label{boundtwo} \Psi'_{n}(L)=0.
\end{eqnarray}

The equation for the mass spectrum has the form
\begin{equation}\label{m-spec}
N_{1}\left(m_{n}\frac{4v_{1}}{\lambda}\right)J_{1}\left(m_{n}\frac{4v_{2}}{\lambda}\right)-
N_{1}\left(m_{n}\frac{4v_{2}}{\lambda}\right)J_{1}\left(m_{n}\frac{4v_{1}}{\lambda}\right)=0.
\end{equation}

Thus we are ready to find the relationship between the
four-dimensional Planck mass and the parameters of the theory.
Taking the first term of equation (\ref{action}), retaining only
the zero mode in (\ref{summ}) and using the fact that
$\Psi_{0}(y)=const$, we can formally get
\begin{equation}
\int \Phi(x,y) R\sqrt{-g}\,d^{4}xdy\supset
\int_{-L}^{L}\left(\frac{\lambda}{4}\,|y|+v_{1}\right)dy \int
R_{(4)}\sqrt{-g}\,d^{4}x
\end{equation}
and
\begin{equation}\label{Planck1}
M^{2}_{Pl}=\int_{-L}^{L}\left(\frac{\lambda}{4}\,|y|+v_{1}\right)dy=\frac{\lambda}{4}L^{2}+2v_{1}L.
\end{equation}
Using (\ref{L}) we get
\begin{equation}\label{Planck2}
M^{2}_{Pl}=\frac{4}{\lambda}\left(v_{2}^2-v_{1}^2\right),
\end{equation}
i.e. the four-dimensional Planck mass is defined by parameters
$\lambda$, $v_{1}$ and $v_{2}$ of the brane tensions. Taking
suitable values of $v_{1}$, $v_{2}$ and $\lambda$ we can solve the
hierarchy problem in the same way as in the ADD model with two
extra dimensions. For example, one can take
$M=\sqrt[4]{\lambda}\sim 100 TeV$ and $L\sim 1 eV^{-1}$ to get
$M_{Pl}\simeq 10^{19} GeV$ (compare with physically interesting
values of parameters in the ADD model with two extra dimensions,
where five-dimensional Planck mass is chosen to be of the order of
$30 TeV$, whereas the size of the extra dimension is chosen to be
of the order of $10^{-2} eV^{-1}$, see \cite{Rub01}). If we
suppose that $v_{1}\sim v_{2}$ (i.e. they are of the same order),
then $M_{v_{i}}\sim\sqrt[3]{v_{i}}$ (where $i=1,2$) should be of
the order of $10^{6}-10^{7}TeV$. It is evident, that this
difference between the energy scales $M$ and $M_{v_{i}}$ appears
because of the new hierarchy between $M$ and $L^{-1}$. One can see
that this hierarchy is inherent to the original ADD scenario with
two extra dimensions too (see also \cite{Rub01}). Of course, the
simplest way to remove hierarchy between $M$ and $M_{v_{i}}$ is to
choose $M$ and $M_{v_{i}}$ ($i=1,2$) to be of the same order,
which results in $L^{-1}\sim M\sim M_{v_{i}}\sim M_{Pl}\simeq
10^{16} TeV$. We can also leave $M$ in the $TeV$ range and
fine-tune parameters $v_{1}$ and $v_{2}$ to make $L^{-1}\simeq
10^{16} TeV$ (in this case $M_{v_{i}}\sim 10^{16} TeV$ also). But
even if only one fundamental parameter of the model is of the
order of $M_{Pl}$, we get the same problem as in ordinary
four-dimensional theory of gravity -- we should explain its
unnaturally large value. In the latter case one should also
explain such extreme fine-tuning between $v_{1}$ and $v_{2}$.
Another possibility to remove this hierarchy is to leave $M$ and
$M_{v_{i}}$ (energy scales of fundamental parameters of the
theory) in the $TeV$ range, but to fine-tune parameters $v_{1}$
and $v_{2}$ to make the size $L$ extremely large
($L\sim\frac{M^{2}_{Pl}}{M^{3}}$), like in the ADD model with one
extra dimension. It is evident that such size of the extra
dimension contradicts observable data.

At the same time the hierarchy between $M\sim 100 TeV$ and $L\sim
eV^{-1}$ is not so dangerous as it seems. We should take into
account that the size of the extra dimension $L$ is not a
fundamental parameter of the model. Indeed, {\em fundamental}
parameters of the model are $\lambda$ and $v_{i}$, whose energy
scales are $M\sim 100 TeV$ and $M_{v_{i}}\sim 10^{6}-10^{7}TeV$,
and the size $L$ is defined by these parameters. One can see that
the largest energy scales in the five-dimensional theory
$M_{v_{i}}\sim 10^{6}-10^{7}TeV$ are much closer to the desirable
value $1 TeV$, which roughly characterizes the energy scale of the
Standard Model on the brane, than the four-dimensional Planck mass
$M_{Pl}\sim 10^{16} TeV$. Moreover, the difference between energy
scales of fundamental parameters $M$, $M_{v_{i}}$ is maximally 5
orders in magnitude, which is very small in comparison with the
hierarchy between the energy scales in ordinary four-dimensional
theory of gravity. We need not to fine-tune parameters $v_{1}$ and
$v_{2}$ -- they should be of the same order only. As for the size
of the extra dimension, its large value appears naturally from
(\ref{L}) and this situation can be interesting in view of
top-table experiments for testing gravity at sub-millimeter
scales.

Now let us discuss how the model can be interpreted. One can see
that at the first sight the model under consideration (with only
one extra dimension) manifests itself through (\ref{Planck1}) as
the ADD model with two extra dimensions (because of the factor
$L^{2}$). At the same time using (\ref{L}) we can rewrite
(\ref{Planck2}) as
\begin{equation}\label{Planck21}
M^{2}_{Pl}=L\left(v_{1}+v_{2}\right),
\end{equation}
which now represents the model as the ADD model with one extra
dimension. Since $\lambda$ and $v_{1}$, $v_{2}$ are all
fundamental parameters, a discrepancy arises. The answer is that
since the size of the extra dimension $L$ is not a fundamental
parameter of the theory, as it was noted above, both
interpretations do not correspond to the real properties of the
model, which is not exactly the ADD model, and therefore should be
rejected.

In the end of this section let us compare the setup discussed
above with some well-known multidimensional models, in which the
scalar field interacting with gravity is used for stabilization of
the extra dimension. The most known model is the one proposed in
\cite{GWise}. But the background solution found in \cite{GWise} is
not an exact one -- backreaction of the scalar field on the metric
is not taken into account in this model. At the same time the
background solution found above is exact, i.e. it satisfies all
equation of motion -- Einstein equations and equation for the
scalar field. Thus, it is more reasonable to compare our setup
with the one proposed in \cite{DeWolfe}, where an exact solution
for gravity and scalar field in five dimensions was obtained (the
so-called stabilized Randall-Sundrum model). It is evident that
consistent comparison can be made only if we transform
(\ref{action}) to the Einstein frame (it can be made with the help
of a conformal rescaling). The bulk part of rescaled action
describes scalar field minimally coupled to five-dimensional
gravity. It differs from the bulk action proposed in
\cite{DeWolfe} in the absence of the bulk potential. Thus, it is
evident that the models have different background solutions.
Nevertheless, methods of stabilization utilized in these models
are similar -- the size of extra dimension is defined by the
boundary conditions on the branes and depends on the parameters of
scalar potentials on the branes, contrary to the case of
\cite{GWise}, where the size of extra dimension is defined by
minimization of the effective four-dimensional scalar potential
(which can be obtained from the averaged five-dimensional action).
In this connection it is necessary to mention paper \cite{Kanti},
in which solution for the system of bulk scalar field without bulk
potential minimally coupled to five-dimensional gravity was found.
Stabilization of the extra dimension is also achieved by the
boundary conditions on the branes. Because of the absence of the
bulk scalar field potential this model seems to be similar in some
sense to the model discussed in this paper. At the same time the
solution found in \cite{Kanti} describes warped brane world and
the dynamics of the scalar field comes from the kinetic term,
contrary to our case with flat bulk and dynamics for the scalar
field coming from non-minimal interaction of this field with
five-dimensional curvature. But since both models are similar with
respect to the absence of the bulk scalar field potential, it
would be interesting to compare predictions of these models in
equal frames.

As for the frame which is chosen for action (\ref{action}), the
background solution in the Jordan frame seems to have the simplest
form and, as it was noted before, it has some "intersections"\
with the ADD model. In addition, the terms in (\ref{action})
corresponding to brane tension have classical form in this frame.
Moreover, since we do not know which frame is the "real"\ one,
there are no any strong objections against choosing Jordan one.

\section{Scalar sector}
\label{sec:2} Now let us make a brief investigation of the scalar
sector of the theory. To this end we need linearized equations of
motion for different components of metric fluctuations and for the
scalar field. Let us denote the fluctuations of metric by
$h_{MN}$, where $M,N=0,..,4$, and fluctuation of the field $\Phi$
by $\varphi$. Thus
\begin{eqnarray}\nonumber
g_{MN}(x,y)=\eta_{MN}+h_{MN}(x,y), \\ \nonumber
\Phi(x,y)=\Phi_{0}+\varphi(x,y),
\end{eqnarray}
where $\eta_{MN}$ is the flat five-dimensional metric. The
corresponding gauge transformations look like
\begin{eqnarray}
h^{(1)}_{MN}=h_{MN}-\partial_{M}\xi_{N}-\partial_{N}\xi_{M}, \\
\varphi^{(1)}=\varphi-\Phi'_{0}\xi_{4}.
\end{eqnarray}
It is not difficult to show, that one can impose the following
gauge on the fields $h_{\mu 4}$, $h_{44}$ and $\varphi$:
\begin{eqnarray}
h_{\mu 4}=0, \\ \label{gaugef}
\varphi(x,y)-\frac{1}{2}\,\Phi_{0}(y)h_{44}(x,y)=f(x),
\end{eqnarray}
i.e. the new field $f$ depends only on four-dimensional
coordinates. Indeed, the corresponding transformations look like
\begin{eqnarray}
\frac{\varphi^{(1)}(x,y)}{\Phi^{2}_{0}}-\frac{h^{(1)}_{44}(x,y)}{2\Phi_{0}}=
\frac{\varphi(x,y)}{\Phi^{2}_{0}}-\frac{h_{44}(x,y)}{2\Phi_{0}}+\left(\frac{\xi_{4}}{\Phi_{0}}\right)',
\end{eqnarray}
which demonstrates the possibility to impose the gauge
(\ref{gaugef}) (see \cite{SV} for details).

We also note that after imposing this gauge we are left with
residual gauge transformations $\xi_{\mu}(x,y)=\epsilon_{\mu}(x)$,
which are responsible for determining the physical degrees of
freedom of the massless four-dimensional graviton.

The corresponding equations of motion can be easily obtained from
(\ref{ee}) and (\ref{scalar}), and for convenience we will present
here expressions for the linear approximation of the term
$\nabla_{M}\nabla_{N}\Phi- g_{MN}g^{AB}\nabla_{A}\nabla_{B}\Phi$:
\begin{eqnarray}\nonumber
\nabla_{\mu}\nabla_{\nu}\Phi-
g_{\mu\nu}g^{AB}\nabla_{A}\nabla_{B}\Phi=\partial_{\mu}\partial_{\nu}\varphi+\frac{1}{2}
h'_{\mu\nu}\Phi'_{0}-h_{\mu\nu}\Phi''_{0}+\\ \label{mu-nu}
+\eta_{\mu\nu}h_{44}\Phi''_{0}
-\eta_{\mu\nu}\Box\varphi-\frac{1}{2}\eta_{\mu\nu}h'\Phi'_{0}-\eta_{\mu\nu}\varphi''+\frac{1}{2}\eta_{\mu\nu}h'_{44}\Phi'_{0},
\end{eqnarray}
\begin{eqnarray}
\nabla_{\mu}\nabla_{4}\Phi- g_{\mu
4}g^{AB}\nabla_{A}\nabla_{B}\Phi=\partial_{\mu}\partial_{4}\varphi-\frac{1}{2}\partial_{\mu}h_{44}\Phi'_{0},
\end{eqnarray}
\begin{eqnarray}
\nabla_{4}\nabla_{4}\Phi- g_{4
4}g^{AB}\nabla_{A}\nabla_{B}\Phi=-\Box\varphi-\frac{1}{2}h'\Phi'_{0}.
\end{eqnarray}

For our purposes we will need the equations for the $\mu 4$- and
$44$-components of metric fluctuations, equation for the field
$\varphi$ and contracted equation for the $\mu\nu$-component of
metric fluctuations. Substituting representation
\begin{eqnarray}\nonumber
h_{\mu\nu}(x,y)&=&b_{\mu\nu}(x,y)-\frac{1}{3}\eta_{\mu\nu}h_{44}(x,y)+\frac{4}{3}\frac{\partial_{\mu}
\partial_{\nu}}{\Box}h_{44}(x,y)-\\ \nonumber
&-&\frac{m_{2}^{2}}{6L}\eta_{\mu\nu}\frac{1}{\Box}\varphi(x,L)+\frac{m_{2}^{2}}{4L}
y^{2}\frac{\partial_{\mu}\partial_{\nu}}{\Box}\varphi(x,L)-\\
&-&\frac{m_{1}^{2}}{6L}\eta_{\mu\nu}\frac{1}{\Box}\varphi(x,0)+\frac{m_{1}^{2}}{4L}
(|y|-L)^{2}\frac{\partial_{\mu}\partial_{\nu}}{\Box}\varphi(x,0)
\end{eqnarray}
into these equations, we get:\\ 1) $\mu 4$-component
\begin{equation}\label{mu-4}
\partial_4 \left(\partial^\nu  b_{\mu\nu}-\partial_\mu b\right)= 0,
\end{equation}
2) $44$-component
\begin{equation}\label{44eq}
\Phi_{0}\left(\partial^\mu \partial^\nu  b_{\mu\nu} - \Box
b\right)-b'\Phi'_{0}+\frac{1}{2L}\left(v_{2}m_{1}^{2}\varphi(0)+v_{1}m_{2}^{2}\varphi(L)\right)-2\Box
f=0,
\end{equation}
3) equation for the field $\varphi$
\begin{equation}\label{varphieq}
\partial^\mu \partial^\nu  b_{\mu\nu} - \Box
b-b''=0,
\end{equation}
where $m_{1}$ and $m_{2}$ -- masses of the scalar field $\varphi$
coming from the stabilizing potentials on the branes,
$\varphi(0)=\varphi(x,0)$, $\varphi(L)=\varphi(x,L)$.

From equations (\ref{mu-4}), (\ref{varphieq}) it follows that
$b''=0$. Using the symmetry conditions one can get $b\,'=0$. With
the help of residual gauge transformations
$\xi_{\mu}(x,y)=\epsilon_{\mu}(x)$ it is possible to impose the
transverse-traceless gauge $b=0$, $\partial^{\mu}b_{\mu\nu}=0$
(see \cite{SV} for details). In this case contracted
$\mu\nu$-equation takes the form
\begin{eqnarray}\label{contrmunu1}\nonumber
\Phi_{0}\left[\frac{3m_{1}^{2}}{2}\varphi(0)\delta(y)+\frac{3m_{2}^{2}}{2}\varphi(L)\delta(y-L)
-\frac{m_{1}^{2}}{L}\varphi(0)-\frac{m_{2}^{2}}{L}\varphi(L)\right]-\\
\label{cmunueq}-2\Phi_{0}\Box
h_{44}-2h''_{44}\Phi_{0}-2h'_{44}\Phi'_{0}=0.
\end{eqnarray}
The boundary conditions can be obtained directly from this
equation. Indeed, let us integrate this equation from $-\epsilon$
to $\epsilon$ and then take the limit $\epsilon\to 0$. Using the
symmetry conditions and the fact that the first derivative of the
fields can have a jump at the point $y=0$, we easily get the
boundary (junction) condition at the point $y=0$. Analogously we
get the boundary (junction) condition at the point $y=L$. Thus the
result looks like
\begin{equation}\label{bound1}
\left[\Phi_{0}\left(\frac{3m_{1}^{2}}{2}\varphi-4h'_{44}\right)\right]|_{y=+0}=0,
\end{equation}
\begin{equation}\label{bound2}
\left[\Phi_{0}\left(\frac{3m_{2}^{2}}{2}\varphi+4h'_{44}\right)\right]|_{y=L-0}=0.
\end{equation}
Thus equations of motion for the scalar degrees of freedom do not
decouple.

It is necessary to note that we use complete equations of motion
in the whole bulk (including contributions with delta-functions
coming from the branes) in the gauge in which branes remain
straight and then extract junction condition at the points the
branes are located at with the help of procedure described above.
Then we solve equations of motion on the segment $[0,L]$ with the
junction conditions playing the role of boundary conditions.
Another approach is utilized, for example, in papers
\cite{Shiromizu,Maeda}, where the induced equation on the brane
and corresponding junction condition (in more general form) are
obtained directly using Gauss and Codazzi equations. Our approach
is more convenient for studying the linearized theory in the whole
bulk, not only in the vicinity of the brane, but it is more
technically complicated than that utilized in
\cite{Shiromizu,Maeda}.

One can see that equation (\ref{contrmunu1}) and corresponding
conditions (\ref{bound1}), (\ref{bound2}) are rather complicated.
In \cite{Csaki} it was suggested to use the "stiff boundary
potential"\ limit for the model proposed in \cite{DeWolfe} to
simplify the analysis. It appears that this limit is very
convenient in our case too. Namely, if $\gamma_{1,2}\to\infty$
(i.e. if $\gamma_{1,2}$ are much larger than other parameters with
the same dimensionality, this also means that $m_{1,2}\to\infty$)
then the conditions for the field $\varphi$ are just
\begin{equation}\label{bound3}
\varphi(0)=\varphi(L)=0,
\end{equation}
i.e. the scalar field $\varphi$ decouples from the theory on the
branes. Thus from equations (\ref{44eq}) and (\ref{cmunueq}) we
get
\begin{equation}\label{eq34}
\Box f=0,
\end{equation}
\begin{eqnarray}\label{scequ}
\Phi_{0}\Box h_{44}+\Phi_{0}h''_{44}+\Phi'_{0}h'_{44}=0.
\end{eqnarray}
Thus (\ref{scequ}) is analogous to equation (\ref{lineq}) for the
tensor modes. In addition we have the following extra boundary
conditions for the field $h_{44}$ coming from (\ref{gaugef}),
(\ref{bound3}) and (\ref{eq34}):
\begin{eqnarray}\label{bound4}
\Box h_{44}|_{y=0}=\Box h_{44}|_{y=L}=0.
\end{eqnarray}
Eigenvalue problem, which corresponds to equations (\ref{scequ})
and (\ref{bound4}), has the form
\begin{eqnarray}\label{firsteq}
\Phi_{0}\mu_{n}^{2}
\tilde\Psi_{n}(y)+\Phi_{0}\tilde\Psi''_{n}(y)+\Phi'_{0}\tilde\Psi'_{n}(y)=0,\\
\label{secondeq} \mu_{n}^{2} \tilde\Psi_{n}(y)|_{y=0}=\mu_{n}^{2}
\tilde\Psi_{n}(y)|_{y=L}=0,
\end{eqnarray}
where $\tilde\Psi_{n}(y)$ is the wave function of the
corresponding four-dimensional mode $\Box
h^{n}_{44}(x)=\mu_{n}^{2}h^{n}_{44}(x)$ with the mass $\mu_{n}$.
Equation (\ref{secondeq}) can be satisfied if:
\begin{enumerate}
\item
$$\mu_{n}\ne 0$$ and $$\tilde\Psi_{n}(0)=\tilde\Psi_{n}(L)=0.$$ But boundary
conditions following from (\ref{firsteq}) are
$\tilde\Psi'_{n}(0)=\tilde\Psi'_{n}(L)=0$ (see (\ref{boundone}),
(\ref{boundtwo})), which are consistent with
$\tilde\Psi_{n}(0)=\tilde\Psi_{n}(L)=0$ only if
$\tilde\Psi_{n}(y)\equiv 0$. \item
$$\mu_{n}=0.$$ In this case equation (\ref{firsteq}) has the
following solution in the bulk
$$
\tilde\Psi'_{n}=\frac{const}{\Phi_{0}}.
$$
From the symmetry conditions for the field $h_{44}$
($h_{44}(-y)=h_{44}(y)$) it follows that $const=0$ and
$\tilde\Psi_{n}$ does not depend on the coordinate of extra
dimension ($\tilde\Psi_{n}=const$). From (\ref{gaugef}) and
(\ref{bound3}) we get
$$\Phi_{0}(0)\tilde\Psi_{n}(0)=\Phi_{0}(L)\tilde\Psi_{n}(L),$$
which implies that $\tilde\Psi_{n}(y)\equiv 0$.
\end{enumerate}
Finally we get $h_{44}(x,y)\equiv 0$. Since $f$ does not depend on
the coordinate of extra dimension, we get $f(x)\equiv 0$. Thus,
the scalar sector totally drops out from the theory, and we get
effective four-dimensional gravity without any extra scalars. The
model appears to be stable in this case. It is evident that
disappearance of scalars is simply an artifact of the "stiff
boundary potential"\ limit used. At the same time this situation
differs considerably from the case of stabilized Randall-Sundrum
model \cite{DeWolfe}, where the radion field survives in the case
of "stiff boundary potential"\ (see \cite{Csaki,BMSV}). At first
glance the absence of the scalar sector in the model seems to be
not very dangerous -- indeed, in paper \cite{SmNPB} it was shown,
that in the Randall-Sundrum model with brane-localized curvature
terms the radion field is absent in the linear approximation
(because of the additional symmetry), whereas the effective theory
on the brane appears to be acceptable. Moreover, the absence of
scalars can ensure the absence of extra effects which could be in
contradiction with the experimental data (such as a very light
radion). At the same time the tensor structure of the
four-dimensional massless graviton seems to be incorrect in the
model as it is. Indeed, the traceless-transverse conditions $b=0$,
$\partial^{\mu}b_{\mu\nu}=0$ following from (\ref{varphieq}), are
not in agreement with the ordinary equation for the massless
graviton
$\Box\left(b_{\mu\nu}-\frac{1}{2}\eta_{\mu\nu}b\right)\sim
T_{\mu\nu}$, where $T_{\mu\nu}$ is the energy-momentum tensor of
matter. In other words, if we place matter on the brane, there
should appear a massless scalar field (this scalar field is not
necessarily consists of the radion and/or field $\varphi$, see
\cite{SmNPB}, where analogous situation arises), which is
additional to the massless graviton as viewed from the point of
view of a four-dimensional observer on the brane, and which
interacts with the trace of the energy-momentum tensor. We can
find an analogy to this situation in electrodynamics. It is a
common knowledge that longitudinal photons do not appear in  the
asymptotic states (on the mass shell), whereas their contribution
is important in the radiative corrections (off the mass shell).
This "reappearing"\ scalar field is very similar to longitudinal
photons: it is absent in the asymptotic states, but it is
absolutely necessary for consistently describing the interaction
off the mass shell. Thus we get the same problem as those in the
DGP model \cite{DGP} and in the theory of four-dimensional massive
gravity \cite{vDV,Zakh}. Maybe it is a consequence of the total
absence of the scalar sector in this approximation in case of
absence of matter on the branes, which "enforces"\ the system to
create an extra field to compensate the absence of the radion and
the field $\varphi$ (for example, in the RS2 model the absence of
the radion field leads even to contradiction \cite{SV}). As for
the general case (arbitrary values of $m_{1}$, $m_{2}$ and
presence of matter on the branes), its analysis is a rather
complicated task. But in this connection it is necessary to note
that there is a lot of possibilities to change the linearized
equations of motion without breakdown of the background solution.
One can add kinetic terms for the field $\Phi$ on the branes,
terms describing non-minimal coupling of the field $\Phi$ to
gravity on the branes and, as a matter of fact, the
four-dimensional scalar curvature:
\begin{equation}
\int_{y=y_{i}}\left(-\alpha_{i}\partial^{\mu}\Phi(x,y_{i})\partial_{\mu}\Phi(x,y_{i})
+ \beta_{i}\Phi(x,y_{i})\tilde R +\psi_{i}\tilde R
\right)\sqrt{-\tilde g}d^{4}x,
\end{equation}
where $\alpha_{i}$, $\beta_{i}$ and $\psi_{i}$ are arbitrary
constants. Such modification, inspired by the scheme proposed in
\cite{DGP}, provides a lot of possibilities to change the spectrum
of the theory considerably (and probably can improve the incorrect
tensor structure of the massless graviton). Moreover, we think
that such modification of original action (\ref{action}) can lead
to more interesting consequences than those for the original
action. But its examination exceeds the limits of this paper.

\section{Conclusion}
\label{concl} In this paper a model describing the scalar field
non-minimally coupled to gravity in the space-time with one extra
dimension is proposed. It possesses remarkable features: branes
{\it with tension} in the {\it flat} five-dimensional background;
a simple and natural way of the extra dimension's size
stabilization; the four-dimensional Planck mass is defined by the
parameters of the scalar field potentials on the branes; the value
of the size of the extra dimension appears to be interesting in
view of testing gravity at sub-millimeter scales.

At the same time it is necessary to make more thorough
investigation of linearized gravity on the branes in general case
(arbitrary values of $m_{1}$ and $m_{2}$) in the presence of
matter on one of the brane (to answer the question about tensor
structure of massless graviton), as well as to obtain a second
variation Lagrangian (to answer the questions about possible
appearance of ghosts in the model). But these problems call for
more detailed investigation.

It should be also noted that just this mechanism is applicable
only in the case of one compact extra dimension. But in any case
this model seems to be quite useful because it suggests a possible
way for constructing models with flat background and tension-full
branes.

\section*{Acknowledgments}
The author is grateful to I. P. Volobuev for valuable discussions.
The work was supported by RFBR grant 04-02-17448, Russian Ministry
of Education and Science grant NS-8122.2006.2 and by the grant for
young scientists MK-8718.2006.2 of the President of Russian
Federation.

\end{document}